\documentclass[twocolumn,showpacs,preprintnumbers,amsmath,amssymb]{revtex4}

\usepackage{graphicx}
\usepackage{dcolumn}
\usepackage{bm}

\begin{document}

\preprint{}

\title{Electric-field control of spin accumulation signals in silicon at room temperature}

\author{Y. Ando,$^{1,2}$ Y. Maeda,$^{1}$ K. Kasahara,$^{1}$ S. Yamada,$^{1}$ K. Masaki,$^{1}$ Y. Hoshi,$^{3}$ K. Sawano,$^{3}$ K. Izunome,$^{4}$ A. Sakai,$^{5}$ M. Miyao,$^{1}$ and K. Hamaya$^{1,6}$\footnote{E-mail: hamaya@ed.kyushu-u.ac.jp}}

\affiliation{$^{1}$Department of Electronics, Kyushu University, 744 Motooka, Fukuoka 819-0395, Japan}%
\affiliation{$^{2}$INAMORI Frontier Research Center, Kyushu University, 744 Motooka, Fukuoka 819-0395, Japan}%
\affiliation{$^{3}$Advanced Research Laboratories, Tokyo City University, 8-15-1 Todoroki, Tokyo 158-0082, Japan}
\affiliation{$^{4}$Principle Technology, Covalent Silicon Corporation, Seiroumachi, Niigata 957-0197, Japan}
\affiliation{$^{5}$Department of Systems Innovation, Osaka University, Toyonaka 560-8531, Japan}
\affiliation{$^{6}$PRESTO, Japan Science and Technology Agency, Sanbancho, Tokyo 102-0075, Japan}%

\date{\today}
\begin{abstract}
We demonstrate spin-accumulation signals controlled by the gate voltage in a metal-oxide-semiconductor field effect transistor structure with a Si channel and a CoFe/$n^{+}$-Si contact at room temperature. Under the application of a back-gate voltage, we clearly observe the three-terminal Hanle-effect signal, i.e., spin-accumulation signal. The magnitude of the spin-accumulation signals can be reduced with increasing the gate voltage. We consider that the gate controlled spin signals are attributed to the change in the carrier density in the Si channel beneath the CoFe/$n^{+}$-Si contact. This study is not only a technological jump for Si-based spintronic applications with gate structures but also reliable evidence for the spin injection into the semiconducting Si channel at room temperature.
\end{abstract}
\maketitle

The progress of silicon-based spintronics (Si spintronics) is splendid in recent years. Many groups have so far demonstrated electrical spin injection and detection through ferromagnet-insulator-Si heterostructures.\cite{Appelbaum,Jonker1,Jonker2,Sasaki1,Jansen,Suzuki} Recently, spin-related phenomena at room temperature were reported in Si-based three- or four-terminal lateral devices.\cite{Jansen,Suzuki,Jonker2} In particular, clear room-temperature spin transport and its manipulation by applying transverse magnetic fields were achieved although the channel used was a heavily doped Si.\cite{Suzuki} 

To date, we have explored the spin injection and detection in Si-based devices without insulators for source and drain contacts in order to reduce the parasitic resistance.\cite{Ando,Ando1} Recently, we studied spin accumulation signals at a ferromagnetic CoFe/$n^{+}$-Si interface by measuring a Hanle effect in three-terminal lateral devices,\cite{Ando1} and found that there is an electrical detectability of the contact for spin accumulation in the Si channel, consistent with the previous work in Fe/GaAs lateral devices.\cite{Lou} If the Hanle-effect signals are arising from the spin accumulation in the Si channel, we should observe the variation in the Hanle-effect signals by changing carrier densities in the Si channel.\cite{Lou,Jonker2} 

In this letter, we demonstrate spin-accumulation signals controlled by the gate voltage in a Si-metal-oxide-semiconductor field effect transistor (MOSFET) structure with a CoFe/$n^{+}$-Si contact at room temperature. Under the application of a back-gate voltage, we clearly observe the three-terminal Hanle-effect signal, i.e., spin-accumulation signal. The magnitude of the spin-accumulation signals can be reduced with increasing the gate voltage. We consider that the gate controlled spin signals can be explained by the change in the carrier density in the Si channel beneath the CoFe/$n^{+}$-Si contact. This is reliable evidence for the spin injection into the semiconducting Si channel at room temperature. From the technological point of view, this study will lead to an acceleration of research and development of Si-based spintronic applications with gate structures.\cite{Sugahara,Dery1}

Ferromagnetic CoFe epitaxial layers with a thickness of $\sim$ 10 nm were grown on (111)-oriented Silicon On Insulator (SOI) by low-temperature molecular beam epitaxy (MBE) at $\sim$ 25 $^{\circ}$C,\cite{Maeda} where the thicknesses of the SOI and buried oxide (BOX) layers are about $\sim$ 75 and 200 nm, respectively, and the carrier density of the SOI layer is $\sim$ 4.5 $\times$ 10$^{15}$cm$^{-3}$ (1 $\sim$ 5 $\Omega$cm) at room temperature. By a combination of the Si epitaxy using an MBE process with an Sb $\delta$-doping technique, an $n$$^{+}$-Si layer (Sb : 1 $\times$ 10$^{19}$cm$^{-3}$) was inserted between CoFe and SOI. Here the Sb $\delta$-doped $n$$^{+}$-Si layer on the channel region was removed by the Ar$^{+}$ ion milling. An ohmic contact (AuSb) for backside heavily doped Si was formed at less than 300 $^{\circ}$C. Conventional processes with electron-beam lithography, Ar$^{+}$ ion milling, and reactive ion etching were used to fabricate three-terminal lateral devices with a backside gate electrode, illustrated in Fig. 1(a). The CoFe/$n^{+}$-Si contact (contact 2) and AuSb ohmic contacts (contact 1 and 3) have lateral dimensions of 1 $\times$ 100 $\mu$m$^{2}$ and 100 $\times$ 100 $\mu$m$^{2}$, respectively. The distance between the contacts 2 and 1 or 3 is $\sim$ 30 $\mu$m. The three-terminal Hanle measurements were performed by a dc method with the current-voltage configuration shown in Fig. 1(a) at room temperature, where a small magnetic field perpendicular to the plane, $B_\text{Z}$, was applied after the magnetic moment of the contact 2 aligned parallel to the plane along the long axis of the contact. 
\begin{figure}
\includegraphics[width=7.5cm]{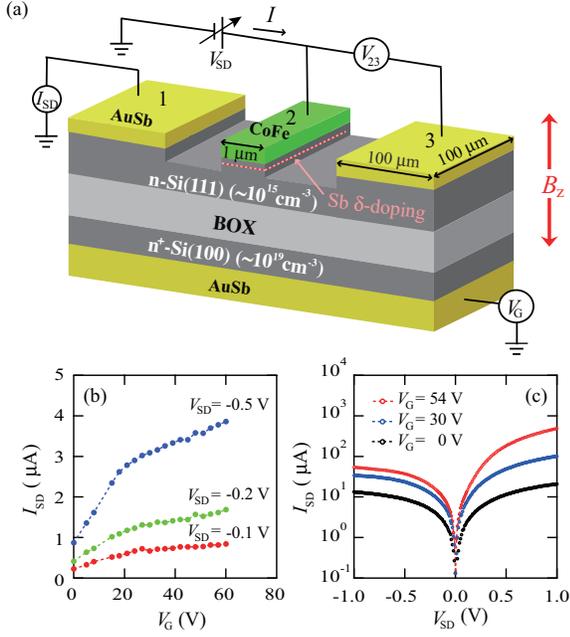}
\caption{(Color online) (a) Schematic diagram of a Si-MOSFET structure with a CoFe/$n^{+}$-Si Schottky-tunnel contact fabricated. (b) $I_\text{SD} -V_\text{G}$ and (c) $I_\text{SD} -V_\text{SD}$ characteristics at room temperature. The constant $V_\text{SD}$ and $V_\text{G}$ values are denoted in each figure. }
\end{figure} 

First, we confirm the operation as a MOSFET for the fabricated device shown in Fig. 1(a). Figure 1(b) shows $I$$_\text{SD}-$$V_\text{G}$ characteristic measured at room temperature at constant bias voltages of $V$$_\text{SD} =$ -0.1, -0.2, and -0.5 V. With increasing $V_\text{G}$, the $I$$_\text{SD}$ value gradually increases for all $V$$_\text{SD}$. This means that the conduction channel is formed from the vicinity of the interface between SOI and BOX. These results clearly indicate that this device can operate as a MOSFET. Because of relatively thick SOI layer, the carrier density of the bulk SOI is still enhanced by further applying $V_\text{G}$. Hence, the $I$$_\text{SD}$ value is not saturated in high $V_\text{G}$ region. Figure 1(c) displays representative $I$$_\text{SD}-$$V_\text{SD}$ characteristics at room temperature at various $V_\text{G}$. Almost symmetric and nonlinear characteristics were observed up to $V_\text{G} =$ 20 V but small asymmetric features with respect to the bias polarity can be seen in $V_\text{G} >$ 20 V. Since these features are markedly different from those resulting from the thermionic emission examined previously in Ref.\cite{Maeda}, tunneling conduction through the CoFe/$n^{+}$-Si interface is dominant factor for the observed $I$$_\text{SD}-$$V_\text{SD}$ characteristics at room temperature. Although we could not make the evident off state at $V_\text{G} =$ 0 V for this device, as shown in Fig. 1(b), there is almost no influence on the main claim of this study.
\begin{figure}
\includegraphics[width=7.5cm]{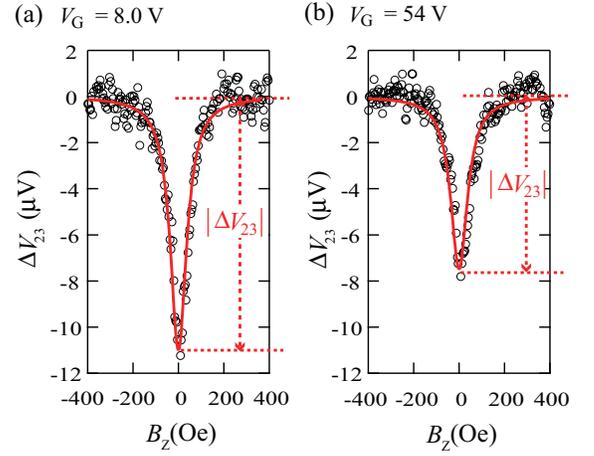}
\caption{(Color online) Room-temperature spin accumulation signals measured at (a) $V_\text{G} =$ 8.0 V and (b) $V_\text{G} =$ 54 V. The applied bias current is a constant value of $I$$_\text{SD} =$ -1.0 $\mu$A, which induces spin accumulation in a Si conduction band by spin injection from CoFe into Si. The red curves are fitting results by the Lorentzian function. }
\end{figure} 

Using this device, we measured the three-terminal voltage, $\Delta$$V_\text{23}$, as a function of $B_\text{Z}$, i.e., Hanle effect. Figure 2(a) shows a $\Delta$$V_\text{23}-$$B_\text{Z}$ curve for $V_\text{G} =$ 8.0 V at $I$$_\text{SD} =$ -1.0 $\mu$A at room temperature, where a quadratic background voltage depending on $B_\text{Z}$ is subtracted from the raw data. Here in this condition ($I$$_\text{SD} <$ 0) the electrons are injected from the spin-polarized states of CoFe into the conduction band of Si. When $B_\text{Z}$ increases from zero to $\pm$200 Oe, a clear voltage change ($|\Delta$$V_\text{23}|$) is observed even at room temperature. The voltage change is caused by the depolarization of the accumulated spins, that is, a Hanle-type spin precession is detected by the three-terminal voltage measurements.\cite{Jansen,Lou,Jonker2,Tran,Sasaki2,Ando1} We could not obtain such Hanle-effect curves in $V_\text{G} < 8.0$ V because the electrical noise was very large. We hereafter concentrate on a constant injection current of $I$$_\text{SD} =$ -1.0 $\mu$A for examining the effect of the application of $V_\text{G}$ ($V_\text{G} \ge 8.0$ V) though the Hanle-effect signals can easily be enhanced by increasing the injection current.\cite{Ando1} We note that the magnitude of $\Delta$$V_\text{23}$, $|\Delta$$V_\text{23}|$, is $\sim$11.5 $\mu$V in Fig. 2(a). Surprisingly, $|\Delta$$V_\text{23}|$ is decreased to $\sim$7.8 $\mu$V when the gate voltage is further applied up to $V_\text{G} =$ 54 V in Fig. 2(b). For both $V_\text{G}$ conditions, a lower limit of spin lifetime ($\tau_\text{S}$) can be obtained using the Lorentzian function, $\Delta$$V_\text{23}$($B_\text{Z}$) $=$ $\Delta$$V_\text{23}(0)$/[1+($\omega_\text{L}$$\tau_\text{S}$)$^{2}$], where $\omega_\text{L} =$ $g\mu_\text{B}$$B_\text{Z}$/$\hbar$ is the Lamor frequency, $g$ is the electron $g$-factor ($g =$ 2), $\mu_\text{B}$ is the Bohr magneton.\cite{Jansen} The fitting results are denoted by the red solid curves in Fig. 2. The $\tau_\text{S}$ values for $V_\text{G} =$ 8.0 and 54 V are estimated to be $\sim$ 1.30 and $\sim$ 1.27 nsec, respectively. It seems that the $\tau_\text{S}$ value is almost constant despite the change in $|\Delta$$V_\text{23}|$. 
\begin{figure}
\includegraphics[width=7.5cm]{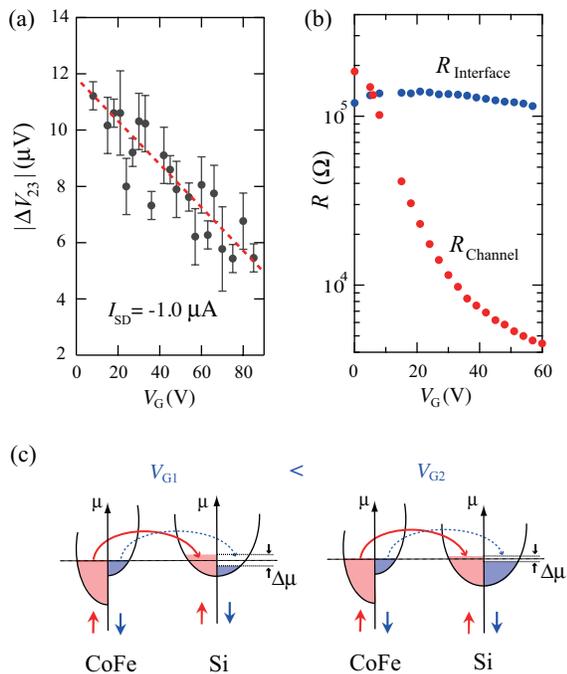}
\caption{(Color online) (a) $|\Delta$$V_\text{23}|$ as a function of $V_\text{G}$ for $I$$_\text{SD} =$ -1.0 $\mu$A at room temperature. (b) The changes in $R_\text{Channel}$ and $R_\text{Interface}$ with increasing $V_\text{G}$ at room temperature. (c) Schematic diagrams of the change in the spin accumulation ($\Delta \mu$) by the application of $V_\text{G}$. }
\end{figure}  

On the basis of the simple spin diffusion model,\cite{Fert,Jedema2,Takahashi,Dery} we consider the observed $|\Delta$$V_\text{23}|$. For three-terminal measurements, the magnitude of the voltage change due to the Hanle-type spin precession can be expressed as follows:\cite{Sasaki2} 
\begin{equation}
\frac{\Delta V}{I} = \frac{P^{2} \lambda_\text{N} \rho_\text{N}}{2A},
\end{equation}
where $P$ is the spin polarization, $\lambda_{\rm N}$ and $\rho_\text{N}$ are the spin diffusion length and resistivity of the nonmagnet used, respectively. $A$ is the contact area. For our fabricated device, when we assume $D \sim$ 40 cm$^{2}$s$^{-1}$ ($n \sim$ 10$^{15}$cm$^{-3}$),\cite{Sze} $\lambda_{\rm Si} \sim$ 2.3 $\mu$m is obtained by using the relationship of $\lambda_{\rm N} =$ $\sqrt{D\tau_\text{S}}$ ($\tau_\text{S} \sim$ 1.3 nsec). Also, $\rho_\text{Si} =$ 1 $\sim$ 5 $\Omega$cm at room temperature is assumed. Since the spin resistance-area-product (spin-$RA$), $\frac{\Delta V_\text{23}}{I_\text{SD}}$$\times$$A$, is obtained to be $\sim$ 1.15 k$\Omega$$\mu$m$^{2}$ for the data in Fig. 2(a), we can roughly obtain 0.14 $<$ $P$ $<$ 0.32 using Eq.(1). The obtained $P$ is consistent with that for CoFe alloys expected.\cite{Monsma} Therefore, our Hanle-effect signals observed here can roughly be considered within the framework of the commonly used diffusion model.\cite{Fert,Jedema2,Takahashi,Dery}  

To discuss the origin of the reduction in $|\Delta$$V_\text{23}|$ in Fig. 2, we explored Hanle-effect signals for various $V_\text{G}$ in detail. Figure 3(a) displays $|\Delta$$V_\text{23}|$ vs $V_\text{G}$ at $I$$_\text{SD} =$ -1.0 $\mu$A at room temperature. With increasing $V_\text{G}$, nearly linear decrease in $|\Delta$$V_\text{23}|$ is obtained. Thus, the observed feature in Fig. 2 is reproduced systematically. Here we also examine the $V_\text{G}$ dependence of channel resistance ($R_\text{Channel}$) and interface resistance ($R_\text{Interface}$) at room temperature, where $R_\text{Channel}$ and $R_\text{Interface}$ can roughly be obtained by using local and nonlocal three-terminal measurements, respectively. As shown in Fig. 3(b), $R_\text{Channel}$ decreases with increasing $V_\text{G}$ while $R_\text{Interface}$ is almost constant irrespective of $V_\text{G}$.\cite{ref} Here we focus on the change in $R_\text{Channel}$ by the application of $V_\text{G}$. Since the thickness of SOI layer is relatively thick ($\sim$ 75 nm), all the changed $R_\text{Channel}$ values after the application of $V_\text{G}$ are not directly associated with the change in $\rho_\text{Si}$ of all the channel regions in this device. However, the application of $V_\text{G}$ should affect the partial change in $\rho_\text{Si}$ of the Si channel at least because $R_\text{Channel}$ does not saturate with increasing $V_\text{G}$. Thus, the carrier density in the Si channel beneath the CoFe/$n^{+}$-Si contact should be enhanced by applying $V_\text{G}$. From these considerations, the monotonous decrease in $|\Delta$$V_\text{23}|$ with applying $V_\text{G}$ can be explained by the decrease in $\rho_\text{N}$ in Eq. (1), where $\rho_\text{N}$ is $\rho_\text{Si}$ of the Si channel beneath the CoFe/$n^{+}$-Si contact.

Phenomenological schematic diagrams of the spin accumulation in the Si channel beneath the CoFe/$n^{+}$-Si contact are shown in Fig. 3(c). Under a condition for spin injection into Si, the spin accumulation ($\Delta \mu$) occurs in the Si conduction band near the quasi Fermi level (left figure). When we increase $V_\text{G}$ from $V_\text{G1}$ to $V_\text{G2}$, the carrier density in the Si conduction channel beneath the CoFe/$n^{+}$-Si contact increases, causing the decrease in $\rho_\text{Si}$. Namely, even if the same $I$$_\text{SD}$ is used for spin injection into the Si channel, tunneling probability of spin-polarized electrons and the density of state in Si at the Fermi level should be varied by the application of $V_\text{G}$, resulting in the reduction in $\Delta \mu$ (right figure). Accordingly, the experimental data can also be explained within a framework based on the simple diffusion model [Eq.(1)]. We note that the present study including gate-induced change in the Hanle-effect signals is exact evidence for the detection of spin-polarized electrons created not in the localized state in the vicinity of the interface but in the Si channel. We convince that the three-terminal Hanle-effect measurements are powerful tool for detection of the spin accumulation in the semiconductor channels. 

In summary, we have demonstrated electric-field control of spin accumulation in Si using a MOSFET structure with a CoFe/$n^{+}$-Si contact. Even at room temperature, we observed clear spin accumulation signals under an application of the gate voltage. The magnitude of the spin accumulation signals was reduced by the increase in the gate voltage. We consider that the reduction in the spin-accumulation signals is attributed to the increase in the carrier density in the Si channel beneath the CoFe/$n^{+}$-Si contact, indicating reliable evidence for the spin injection into the semiconducting Si channel at room temperature. This study also includes a technological jump for Si-based spintronic applications with gate structures. 

This work was partly supported by Precursory Research for Embryonic Science and Technology (PRESTO) from Japan Science and Technology Agency, and Semiconductor Technology Academic Research Center (STARC). Three of the authors (Y.A. K.K. and S.Y.) acknowledge Japan Society for the Promotion of Science (JSPS) Research Fellowships for Young Scientists.



\begin{thebibliography}{11}
\bibitem{ref}
From this data, we can estimate the tunnel resistance of the CoFe/$n^{+}$-Si interface, $R_\text{Interface} \times A$, to be $\sim$ 10$^{7}$ $\Omega$$\mu$m$^{2}$.
\end{thebibliography}
\end{document}